\documentclass[aps,pre,twocolumn,groupedaddress,showpacs,amsmath,floatfix]{revtex4}
    \usepackage{bm}
    \usepackage{graphicx}
\newcommand\beq{\begin{equation}}
\newcommand\eeq{\end{equation}}
\newcommand\beqa{\begin{eqnarray}}
\newcommand\eeqa{\end{eqnarray}}
\begin{document}
%\preprint{}
\title{Equation of state of additive hard-disk fluid mixtures:\\
 A critical analysis of two recent proposals}
\author{M. L\'{o}pez de Haro}
\email{malopez@servidor.unam.mx}
\affiliation{Centro de Investigaci\'on en Energ\'{\i}a, UNAM, Temixco,
Morelos 62580, Mexico}
\author{S. B. Yuste}
\email{santos@unex.es}
\author{A. Santos}
\email{andres@unex.es}
\affiliation{Departamento de F\'{\i}sica, Universidad de Extremadura,
E-06071 Badajoz, Spain}
%\thanks{}
\date{\today}
\begin{abstract}
A detailed analysis of two different theoretical equations of state for
a binary mixture of additive hard disks [C.\ Barrio and J.\ R.\ Solana,
Phys.\
Rev.\ E {\bf 63}, 011201 (2001); A.\ Santos, S.\ B.\ Yuste and M.\
L\'{o}pez
de Haro, Mol.\ Phys.\ {\bf 96}, 1 (1999)], including their
comparison with
Monte Carlo results, is carried out. It is found that both
proposals, which
require the equation of state of the single component
system as input, lead
to comparable accuracy  {when the same input is
used in both}, but the
one advocated by  {Santos \textit{et al.}} is simpler and complies 
with the
exact limit in which the small disks are point particles.
\end{abstract}
%\keywords{}
\maketitle
%\section{Introduction\label{sec1}}
\section{Introduction\label{sec1}}
%\subsection{}
%\subsubsection{}
Despite being in principle a simpler system, hard-disk fluid
mixtures have received much less attention in the literature than
fluid mixtures of hard spheres. This may well be tied to the fact that
up to now no analytical solution to the Percus--Yevick equation
has been found for even dimensionality. In any case, what this has
meant is that fewer results are available for fluid mixtures of
hard disks than for hard-sphere mixtures. In particular, a very
scarce number of proposals for the equation of state (EOS) of
these mixtures has been made \cite{BXHB88,W98,SYH99,S99,BS01},
although the trend seems to be reversing recently, and even fewer
simulations have been performed to assess the value of such
proposals.
In a recent paper, Barrio and Solana \cite{BS01} proposed an EOS
for a binary mixture of additive hard disks. Such an equation
reproduces the (known) exact second and third virial coefficients
of the mixture and may be expressed in terms of the EOS of a
single component system. They also performed Monte Carlo
simulations and found that their recipe was very accurate provided
an  {also very} accurate EOS for the single component system (in their 
case it
was the EOS proposed by Woodcock \cite{W76}) was taken as input.
The comparison with other EOS for the mixture available in the
literature indicated that their proposal does the best job with
respect to the Monte Carlo data. Among these other EOS for the
binary mixture  {considered in Ref.\ \cite{BS01}}, only the one 
introduced by  {Santos \textit{et al.}} a few years
ago \cite{SYH99} also shares with Barrio and Solana's EOS the fact
that it may be expressed in terms of the EOS for a single
component system. The aim of the present  {paper} is to
present a detailed analysis of these two different equations of
state since the comparison made in Ref.\ \cite{BS01} may be
misleading in that it was not performed by taking the {\em same}
EOS for the single component system in both proposals. A
preliminary report of this work can be found in Ref.\
\cite{HYS01}.

In order to carry out the analysis, the paper is organized as
follows. In Sec.\ \ref{sec2} we recall the two different
formulations for the EOS of a binary mixture of additive hard
disks in a unified notation as well as provide the explicit
(approximate) expressions for the EOS of the single component
system that will be used in the actual calculations. This is
followed in Sec.\ \ref{sec3} by a discussion of the results and
some concluding remarks.
\section{The equation of state of a binary
mixture of additive hard disks\label{sec2}}
Let us consider a binary mixture of additive hard disks of
diameters $\sigma_{1}$ and $\sigma _{2}$. The total number density
is $\rho $, the mole fractions are $x_{1}$ and $x_{2}=1-x_{1}$,
and the packing fraction is $\eta =\frac{\pi }{4}\rho \langle
\sigma ^{2}\rangle $, where $\langle \sigma ^{n}\rangle \equiv
\sum_{i}x_{i}\sigma _{i}^{n}$. Let $Z=p/\rho k_{B}T$ denote
the compressibility factor, $p$ being the pressure, $T$ the
absolute temperature and $k_{B}$ the Boltzmann constant. Then,
Barrio and Solana's EOS for a binary mixture of hard disks,
$Z_{\text{m}}^{\text{BS}}(\eta )$,
may be written in terms of a given EOS for a single component system, 
$Z_{\text{s}}(\eta )$, as
\begin{equation}
Z_{\text{m}}^{\text{BS}}(\eta )=1+\frac{1}{2}(1+\beta \eta )\left(
1+\xi \right) \left[ Z_{\text{s}}(\eta )-1\right] ,  \label{1}
\end{equation}
where $\xi \equiv \langle \sigma \rangle ^{2}/\langle \sigma
^{2}\rangle $ and $\beta $ is adjusted as to reproduce the exact
third virial coefficient for the mixture $B_{3}$, namely
\begin{equation}
\beta =\frac{B_{3}}{(\pi /4)^{2}\langle \sigma ^{2}\rangle
^{2}(1+\xi )}- \frac{b_{3}}{2}.  \label{2}
\end{equation}
Here, $b_{3}=4(4/3-\sqrt{3}/\pi )$ is the reduced third virial
coefficient for the single component system  {and} $B_{3}$ is given
by \cite{BXHB88}
\begin{eqnarray}
B_{3}&=&\frac{\pi }{3}\left( a_{11}x_{1}^{3}\sigma
_{1}^{4}+3a_{12}x_{1}^{2}x_{2}\sigma
_{12}^{4}\right.\nonumber\\
&&\left.+3a_{21}x_{1}x_{2}^{2}\sigma
_{12}^{4}+a_{22}x_{2}^{3}\sigma _{2}^{4}\right) ,  \label{3}
\end{eqnarray}
where
\begin{eqnarray}
a_{ij}&=&\pi +2\left( \sigma _{i}^{2}/\sigma _{ij}^{2}-1\right) \cos
^{-1}(\sigma _{i}/2\sigma _{ij})\nonumber\\
&&-\left(4\sigma _{ij}^{2}/\sigma _{i}^{2}-1\right)^{1/2}
(1+\sigma _{i}^{2}/2\sigma _{ij}^{2})\sigma _{i}^{2}/2\sigma
_{ij}^{2} \label{4}
\end{eqnarray}
and $\sigma _{ij}=(\sigma _{i}+\sigma _{j})/{2}$.

The EOS for the mixture, consistent with a given EOS for a single
component system,  introduced recently by Santos \textit{et al.} reads 
\cite{SYH99}
\begin{equation}
Z_{\text{m}}^{\text{SYH}}(\eta )=\left( 1-\xi \right)
\frac{1}{1-\eta }+\xi Z_{\text{s}}(\eta ). \label{5}
\end{equation}
We stress the fact that Eq.\ (\ref{5}) is simpler than Eq.\
(\ref{1}) [which must be complemented with Eqs.\
(\ref{2})--(\ref{4})]. In addition, the structure of Eq.\
(\ref{5}) is valid for any number of components, while  {Eq.\ 
(\ref{1}) requires} the third
virial coefficient,  {which} is known exactly only for \textit{binary} 
mixtures.
In order to proceed with a  {quantitative} analysis of Eqs.\ 
(\ref{1}) 
 and
(\ref{5}), we have to specify  $Z_{\text{s}}(\eta )$. While many
choices are available, we will restrict ourselves to the three
following EOS of the single component system:
\begin{enumerate}
\item[a)] Woodcock's EOS \cite{W76}
\begin{equation}
Z_{\text{s}}^{\text{W}}(\eta )=\frac{1+3\eta /\eta _{0}}{1-\eta /\eta _{0}}%
+\sum_{n=2}^{6}\left( b_{n}\eta _{0}^{n-1}-4\right) \left( \eta
/\eta _{0}\right) ^{n-1},  \label{Woodcock}
\end{equation}
where $\eta _{0}=\left( \sqrt{3}/6\right) \pi $ is the value of
the crystalline close-packing and the $b_{n}$ $(n=2,\ldots,6)$ are the
(known) reduced virial coefficients \cite{JvR93}.
\item[b)] The Levin  approximant of Erpenbeck and Luban \cite{EL85}
\begin{equation}
Z_{\text{s}}^{\text{EL}}(\eta )=\frac{\sum\limits_{n=0}^{4}p_{n}\eta ^{n}}{
\sum\limits_{n=0}^{5}q_{n}\eta ^{n}},  \label{Levin}
\end{equation}
where $q_{n}=(-1)^{n}\left(
\begin{tabular}{l}
$6$ \\
$n$
\end{tabular}
\right) \left( 1-n/6\right) ^{5}b_{6}/b_{6-n}$ and 
$p_{n}=\sum_{m=0}^{n}b_{n+1-m}q_{m}$.
\item[c)] The  EOS proposed by Santos \textit{et al.} \cite{SHY95,HSY98}
\begin{equation}
Z_{\text{s}}^{\text{SHY}}(\eta )=\left( 1-2\eta +\frac{2\eta
_{0}-1}{\eta _{0}^{2}}\eta ^{2}\right) ^{-1}.  \label{SHY}
\end{equation}
\end{enumerate}
 {The EOS (\ref{Woodcock}) and (\ref{Levin}) are more complex than 
the EOS (\ref{SHY}) in the sense that they require the exact knowledge of 
the first six virial coefficients, while Eq.\ (\ref{SHY}) is constructed by 
using the first two virial coefficients only and enforcing a pole at 
$\eta=\eta_0$.
Despite its simplicity, however, Eq.\ (\ref{SHY}) does a remarkably good job 
when compared with simulation data, although it is of course 
less accurate than the more sophisticated EOS (\ref{Woodcock}) and 
(\ref{Levin}) \cite{SHY95}.}

\section{Discussion\label{sec3}}
\begin{figure}[tbp]
\includegraphics[width=.90 \columnwidth]{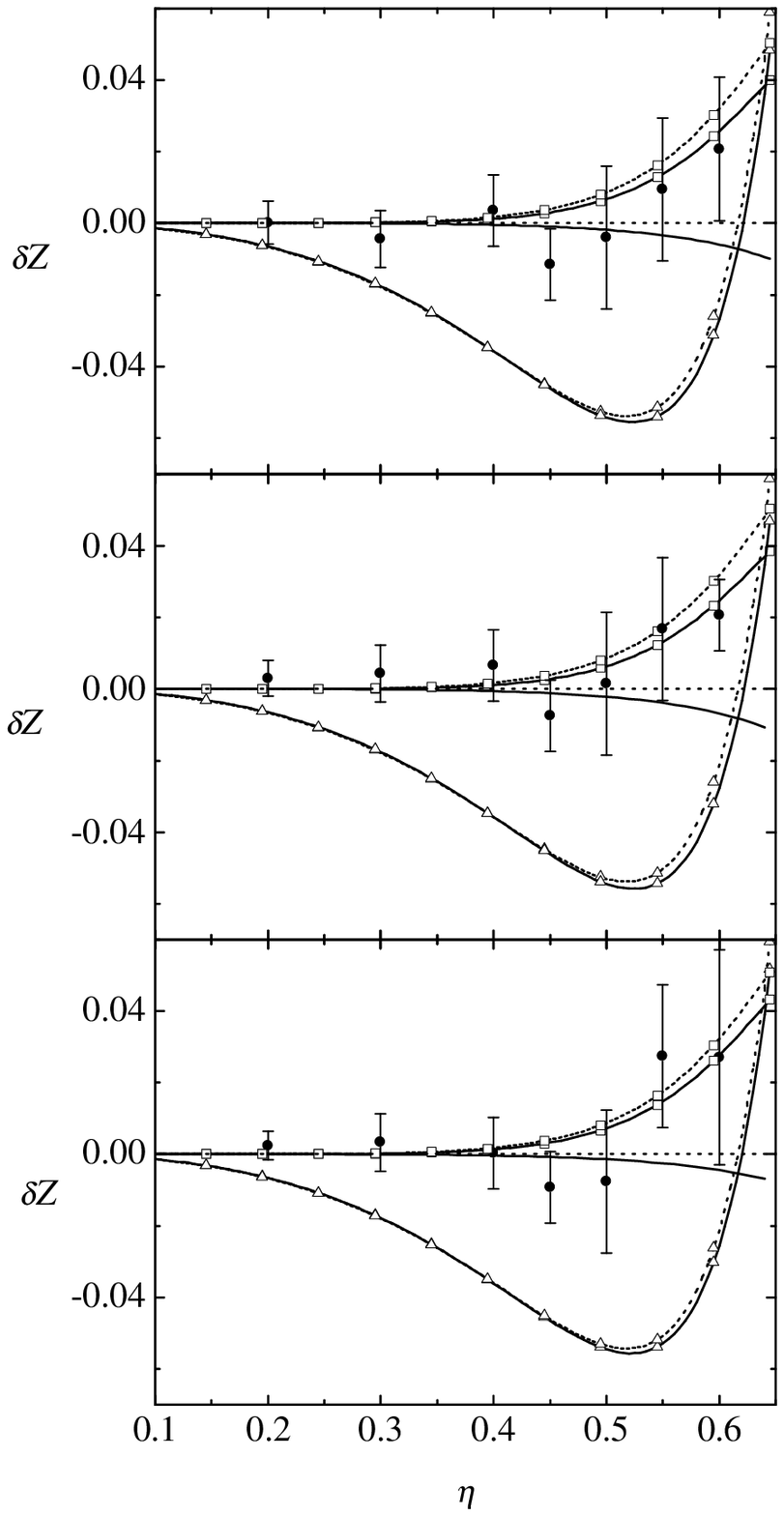}
\caption{ {Plot of the difference 
$\delta Z(\eta)\equiv Z_{\text{m}}(\eta)-Z_{\text{m}}^{\text{BS(W)}}(\eta)$ 
for a size ratio $\sigma_2/\sigma_1=2/3$ and for $x_1 =0.25$ (top panel), 
$x_1 =0.50$ (middle panel), and $x_1 =0.75$ (bottom panel). The 
filled circles are Monte Carlo data \protect\cite{BS01}, the dashed lines 
refer to the proposal (\protect\ref{1}), and the solid lines refer to the 
proposal (\protect\ref{5}). The EOS of the single component fluid, 
$Z_{\text{s}}(\eta)$, used as input are Eq.\ (\protect\ref{Woodcock}) (lines 
without symbols), Eq.\ (\protect\ref{Levin}) (lines 
with squares), and Eq.\ (\protect\ref{SHY}) (lines 
with triangles)}.\label{fig1}}
\end{figure}

\begin{figure}[tbp]
\includegraphics[width=.90 \columnwidth]{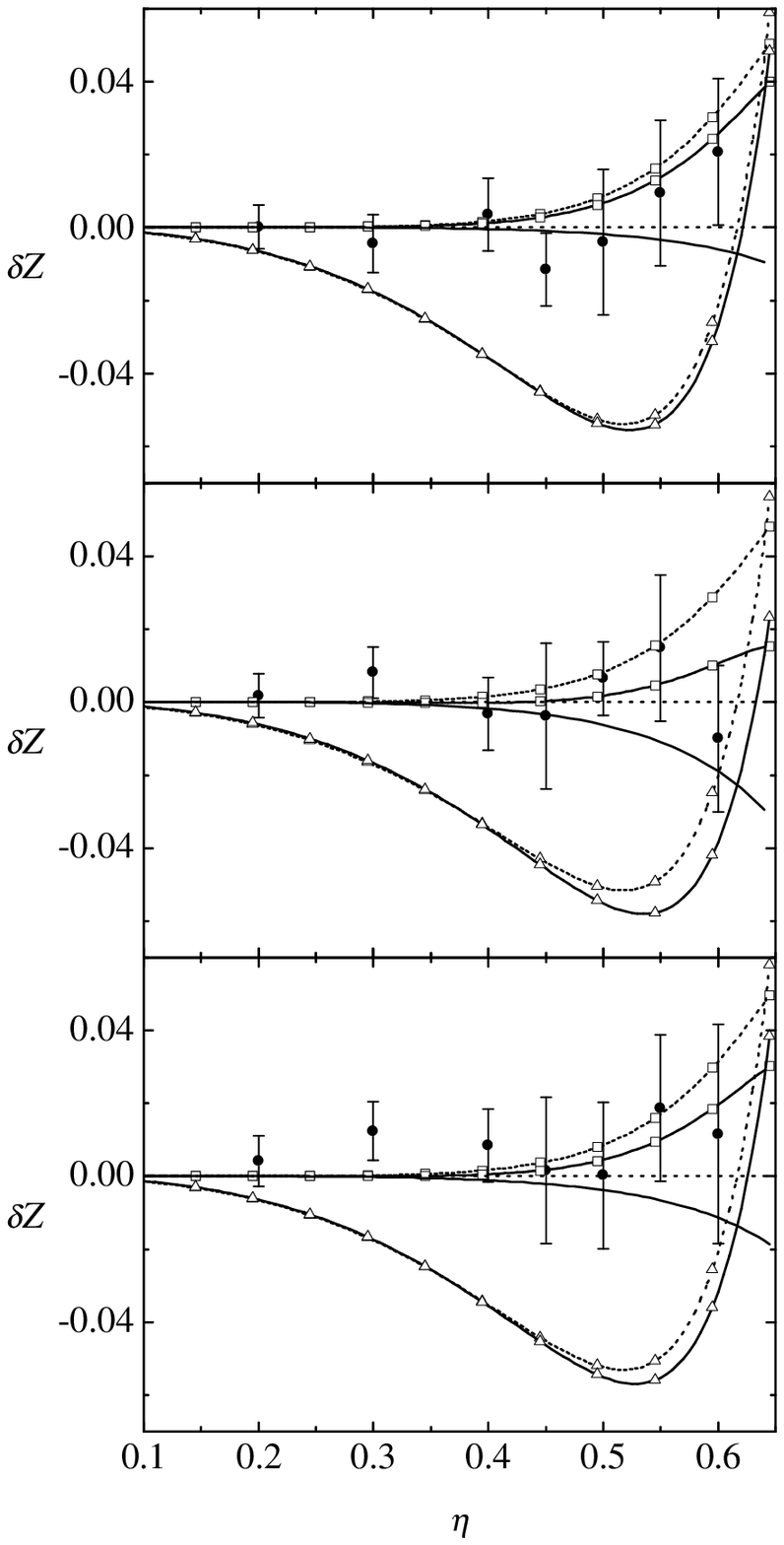}
\caption{ {Same as in Fig.\ \protect\ref{fig1}, but  for a size 
ratio $\sigma_2/\sigma_1=1/2$}.\label{fig2}}
\end{figure}

\begin{figure}[tbp]
\includegraphics[width=.90 \columnwidth]{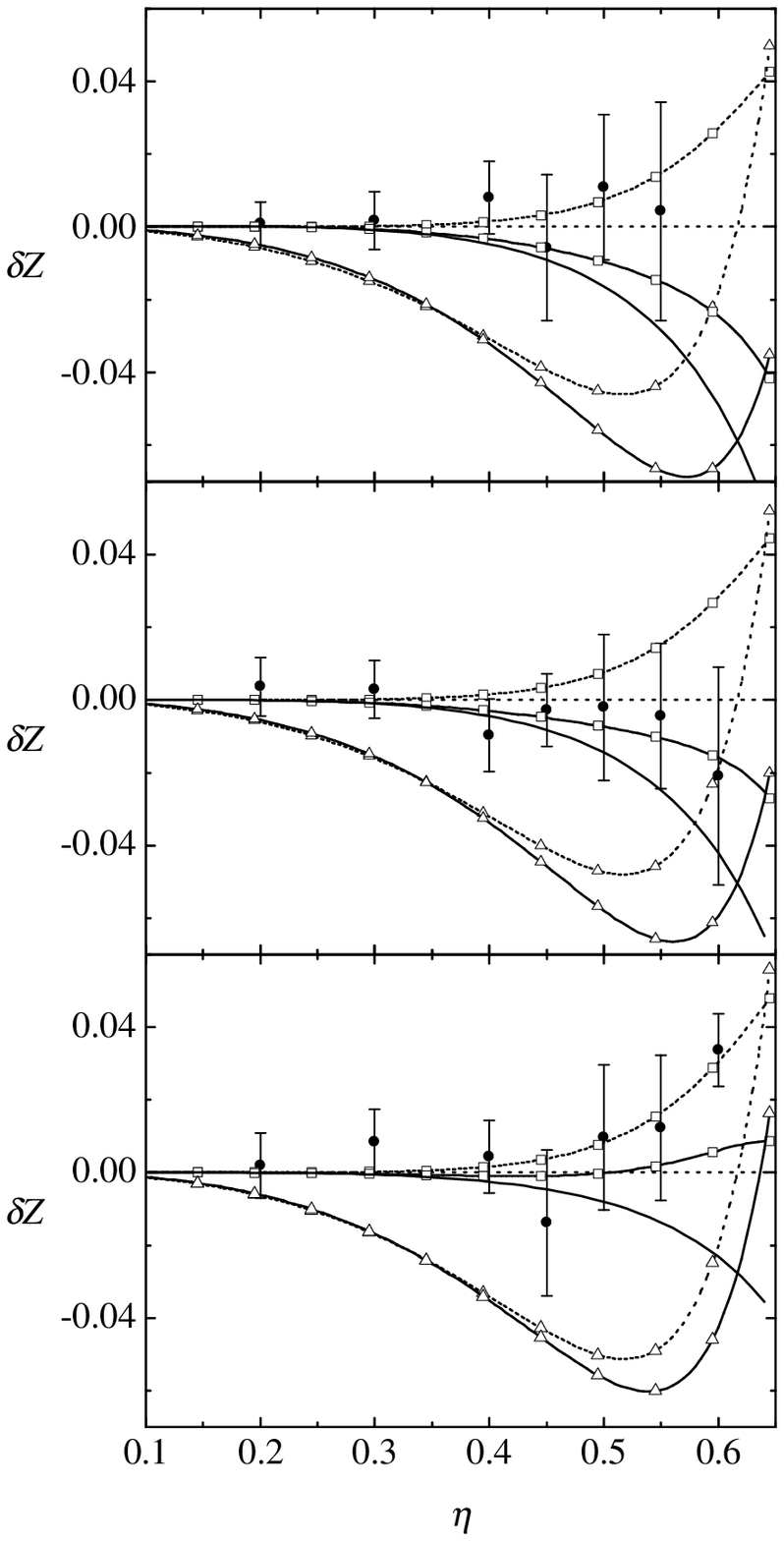}
\caption{ {Same as in Fig.\ \protect\ref{fig1}, but  for a size 
ratio $\sigma_2/\sigma_1=1/3$}.\label{fig3}}
\end{figure}

In Table \ref{table1}, we show the results of Eqs.\ (\ref{1}) and
(\ref{5}) when Woodcock's EOS \cite{W76}, Eq.\ (\ref{Woodcock}), for
the single component system is used as input in both  {equations}, as 
well as
the available MC data \cite{BS01}. As seen in Table \ref{table1}, it is fair 
to
say that both recipes are of comparable accuracy with respect to
the Monte Carlo results, their difference being generally smaller
than the error bars of the simulation data, although
$Z_{\text{m}}^{\text{BS}}(\eta )$ performs slightly better than
$Z_{\text{m}}^{\text{SYH}}(\eta )$. This may be fortuitous since
if one
takes for $Z_{\text{s}}(\eta )$   in
Eqs.\ (\ref{1}) and (\ref{5}) the Levin  approximant \cite{EL85}, Eq.\ 
(\ref{Levin}), (which is known to give the most accurate
approximation to the single component compressibility
factor \cite{EL85,SHY95}) the apparent (slight) superiority of 
$Z_{\text{m}}^{\text{BS}}(\eta )$ is no longer there. For instance, the
theoretical values of Table \ref{table1} corresponding to the
packing fraction $\eta =0.6 $ are increased by about 0.03 when the
Levin  approximant rather than
Woodcock's EOS is used as input, so that in this case the accuracy of 
$Z_{\text{m}}^{\text{SYH}}(\eta )$ is  {generally} slightly better 
than that of $Z_{\text{m}}^{\text{BS}}(\eta )$. This is further illustrated 
in
Figs.\ \ref{fig1}-\ref{fig3}, where in order to enhance the differences we 
have plotted $Z_{\text{m}}(\eta 
)-Z_{\text{m}}^{\text{BS(W)}}(\eta )$ versus $\eta $, 
$Z_{\text{m}}^{\text{BS(W)}}(\eta )$ indicating the use of Eq.\
(\ref{1}) taking for $Z_{\text{s}}(\eta )$ the EOS by Woodcock.
%and the X labeling the different choices.
 {The figures show that, in general, the differences between the EOS 
(\ref{1}) and the EOS (\ref{5}), taking of course the same 
$Z_{\text{s}}$ as input, are smaller than or of the order of the error bars 
of the simulation data. As expected, a better agreement with the simulation 
data is obtained  when either of the more accurate EOS 
(\ref{Woodcock}) or (\ref{Levin})  is used as input instead of the 
much simpler EOS (\ref{SHY}).
It is interesting to remark that the best agreement at the two largest 
densities, $\eta=0.55$ and $\eta=0.6$, corresponds to the use of the Levin 
approximant for $Z_{\text{s}}$.} 
\begin{table*}[tbp]
\caption{Compressibility factor for different binary mixtures of
hard disks as obtained from Monte Carlo simulations, from Eq.\
({\ref{1}}), and from Eq.\ ({\ref{5}}). In the two latter,
Woodcock's equation of state for the single component system is
used. \label{table1}}
\begin{ruledtabular}
\begin{tabular}{ccccccccccc}
&  & \multicolumn{3}{c}{$x_1=0.25$} &
\multicolumn{3}{c}{$x_1=0.50$} & \multicolumn{3}{c}{$x_1=0.75$} \\
\cline{3-5}\cline{6-8}\cline{9-11} $\sigma_2/\sigma_1$ & $\eta$ &
MC\footnotemark[1]&
Eq.\ (\ref{1}) & Eq.\ (\ref{5}) & MC\footnotemark[1]& Eq.\ (\ref{1}) & Eq.\ 
(\ref{5}) & MC\footnotemark[1]& Eq.\ (\ref{1}) & Eq.\ (\ref{5}) \\
\colrule
$2/3$ & 0.20 & 1.559(6) & 1.559 & 1.559 & 1.561(5) &
1.558 & 1.558 & 1.565(4) & 1.563 & 1.563 \\ & 0.30 & 2.036(8) &
2.040 & 2.040 & 2.043(8) & 2.039 & 2.039 & 2.051(8) & 2.048 &
2.048 \\ & 0.40 & 2.79(1) & 2.79 & 2.79 & 2.79(1) & 2.78 & 2.78 &
2.80(1) & 2.80 & 2.80 \\ & 0.45 & 3.31(1) & 3.32 & 3.32 & 3.31(2)
& 3.32 & 3.32 & 3.33(1) & 3.34 & 3.34 \\ & 0.50 & 4.02(2) & 4.02 &
4.02 & 4.02(1) & 4.02 & 4.02 & 4.04(2) & 4.05 & 4.05 \\ & 0.55 &
4.98(2) & 4.97 & 4.97 & 4.98(2) & 4.96 & 4.96 & 5.03(2) & 5.00 &
5.00 \\ & 0.60 & 6.31(2) & 6.29 & 6.28 & 6.30(1) & 6.28 & 6.27 &
6.36(3) & 6.33 & 6.33 \\ &  &  &  &  &  &  &  &  &  &  \\ $1/2$ &
0.20 & 1.534(6) & 1.536 & 1.536 & 1.540(6) & 1.538 & 1.538 &
1.556(7) & 1.552 & 1.552 \\ & 0.30 & 1.998(7) & 1.995 & 1.995 &
2.008(7) & 2.000 & 2.000 & 2.039(8) & 2.027 & 2.026 \\ & 0.40 &
2.72(1) & 2.70 & 2.70 & 2.71(1) & 2.71 & 2.71 & 2.77(1) & 2.76 &
2.76 \\ & 0.45 & 3.20(2) & 3.21 & 3.21 & 3.22(2) & 3.22 & 3.22 &
3.29(2) & 3.29 & 3.29 \\ & 0.50 & 3.88(1) & 3.88 & 3.87 & 3.90(2)
& 3.89 & 3.89 & 3.98(2) & 3.98 & 3.98 \\ & 0.55 & 4.79(2) & 4.77 &
4.76 & 4.81(2) & 4.80 & 4.78 & 4.93(2) & 4.91 & 4.90 \\ & 0.60 &
6.03(3) & 6.02 & 6.00 & 6.04(2) & 6.05 & 6.03 & 6.22(3) & 6.21 &
6.20 \\ &  &  &  &  &  &  &  &  &  &  \\ $1/3$ & 0.20 & 1.491(6) &
1.490 & 1.490 & 1.510(8) & 1.506 & 1.506 & 1.538(9) & 1.536 &
1.536 \\ & 0.30 & 1.907(8) & 1.905 & 1.904 & 1.940(8) & 1.937 &
1.936 & 2.004(9) & 1.996 & 1.995 \\ & 0.40 & 2.55(1) & 2.54 & 2.54
& 2.59(1) & 2.60 & 2.60 & 2.71(1) & 2.71 & 2.70 \\ & 0.45 &
2.99(2) & 3.00 & 2.99 & 3.07(1) & 3.07 & 3.06 & 3.20(2) & 3.21 &
3.21 \\ & 0.50 & 3.60(2) & 3.59 & 3.57 & 3.69(2) & 3.69 & 3.68 &
3.89(2) & 3.88 & 3.87 \\ & 0.55 & 4.39(3) & 4.39 & 4.36 & 4.52(2)
& 4.52 & 4.50 & 4.79(2) & 4.78 & 4.76 \\ & 0.60 &  & 5.49 & 5.44 &
5.66(6) & 5.68 & 5.64 & 6.06(1) & 6.03 & 6.00
\end{tabular}
\footnotetext[1]{Ref.\ \protect\cite{BS01}} 
\end{ruledtabular}
\end{table*}
\begin{figure}[tbp]
\includegraphics[width=0.90 \columnwidth]{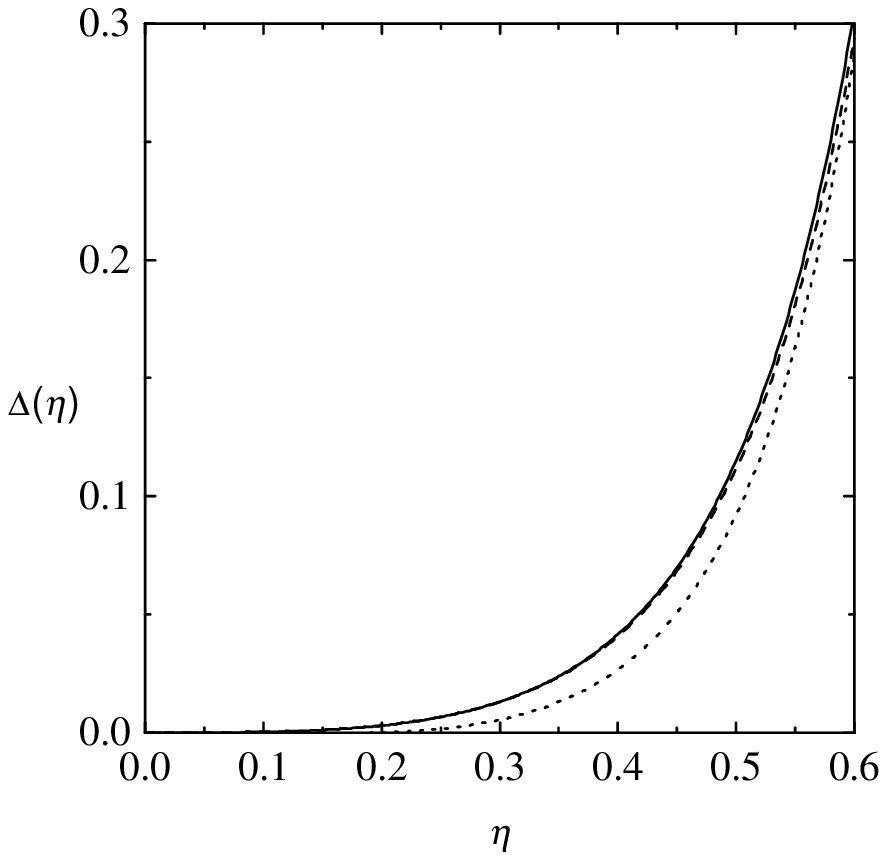}
\caption{Plot of $\Delta (\protect\eta )$, Eq.\ (\protect\ref{8}), by 
assuming  Woodcock's EOS (dashed line),  the Levin  approximant
(solid line), and the SHY  EOS
(dotted line) for the pure fluid. }
\label{fig4}
\end{figure}

Let us try to understand why both EOS  {for the mixture} give 
practically equivalent
results  {when the same input is used in both}. First, it may be shown 
that
$Z_{\text{m}}^{\text{SYH}}(\eta )$, while not reproducing the
exact third virial coefficient $B_{3}$, yields a very good
estimate of it \cite{SYH01},  {namely $B_3\simeq 
\left[1+(b_3-1)\xi\right](\pi/4)^2\langle \sigma^2\rangle^2$}. If we replace 
that estimate into Eq.\
(\ref{2}), we get
\begin{equation}
\beta \simeq \frac{1-\xi }{1+\xi }\left( 1-\frac{b_{3}}{2}\right).
\label{6}
\end{equation}
By using this estimate in Barrio and Solana's EOS, Eq.~(\ref{1}), we have
\begin{equation}
Z_{\text{m}}^{\text{BS}}(\eta )-Z_{\text{m}}^{\text{SYH}}(\eta
)\simeq
\left( 1-\xi \right) \Delta (\eta ),  \label{7}
\end{equation}
where
\begin{equation}
\Delta (\eta )\equiv \frac{1}{2}\left[ 1-\left(\frac{b_{3}}{2}-1\right)
\eta \right] \left[ Z_{\text{s}}(\eta )-1\right] -\frac{\eta }{1-\eta }.
\label{8}
\end{equation}
According to the approximation involved in Eq. (\ref{7}), the difference 
$Z_{\text{m}}^{\text{BS}}(\eta )-Z_{\text{m}}^{\text{SYH}}(\eta )$ is small 
if
the asymmetry of the mixture is small ($\xi \lesssim 1$) and/or $\Delta
(\eta )$ is small. The function $\Delta (\eta )$ is plotted in Fig.\ 
\ref{fig4} for the cases where $Z_{\text{s}}(\eta )$ is given by  Woodcock's 
EOS, by the Levin  approximant, and by the SHY EOS. In
all instances it is practically zero up to $\eta \approx 0.2$ but then it
grows rapidly. The most disparate mixture considered in Barrio and Solana's
simulations corresponds to $x_{1}=0.25$, $\alpha \equiv \sigma _{2}/\sigma
_{1}=1/3$, which yields $\xi =0.75$. This explains the fact that 
$Z_{\text{m}}^{\text{BS}}(\eta )-Z_{\text{m}}^{\text{SYH}}(\eta )\lesssim 
0.05$ in the
simulated cases. It should be noted however that Eq.\ (\ref{7}) tends to
overestimate the actual difference $Z_{\text{m}}^{\text{BS}}(\eta 
)-Z_{\text{m}}^{\text{SYH}}(\eta )$ (for instance, in the case where 
$x_{1}=0.25$, $\alpha =1/3$, $\eta =0.6$ and $Z_{s}(\eta )$ is given by 
Woodcock's EOS,
this difference is $0.05$, while the prediction of Eq.\ (\ref{7}) yields  
$0.07$), so that its main purpose is to illustrate the fact
that both EOS yield practically equivalent results for not very asymmetric
mixtures. On the other hand, more important differences can be expected for
disparate mixtures, especially in the case of large densities. At a given
density and a given diameter ratio $\alpha \leq 1$ the smallest value of the
parameter $\xi $ corresponds to a mole fraction $x_{1}=\alpha /(1+\alpha )$
for the large disks, namely $\xi =4\alpha /(1+\alpha )^{2}$. Thus, $\xi \ll 1
$ if $\alpha \ll 1$ and, according to Eq.\ (\ref{7}), 
$Z_{\text{m}}^{\text{BS}}(\eta )-Z_{\text{m}}^{\text{SYH}}(\eta )\simeq 
\Delta (\eta )$.

Let us consider now the limit in which the small disks become point
particles ($\alpha \rightarrow 0$) and occupy a negligible fraction of the
total area. In that case, the compressibility factor of the mixture must 
reduce to \cite{SYH99,H96}
\begin{equation}
Z_{\text{m}}(\eta )\rightarrow \frac{x_{2}}{1-\eta }+x_{1}Z_{\text{s}}(\eta
).  \label{9}
\end{equation}
The first term represents the (ideal gas) partial pressure due to the point
particles in the available area (i.e., the total area minus the area
occupied by the large disks), while the second term represents the partial
pressure associated with the large disks. In the limit $\alpha \rightarrow 0$
with $x_{1}$ finite (or, more generally,  {for} $\alpha\ll x_{1}$), 
we have $\xi
\rightarrow x_{1}$ and $\beta \rightarrow (1-b_{3}/2)x_{2}/(1+x_{1})$ [note
 {that} in  {this} limit the approximation (\ref{6}) becomes 
correct], so that
\begin{equation}
Z_{\text{m}}^{\text{SYH}}(\eta )\rightarrow \frac{x_{2}}{1-\eta }+x_{1}Z_{%
\text{s}}(\eta ),  \label{10}
\end{equation}
\begin{equation}
Z_{\text{m}}^{\text{BS}}(\eta )\rightarrow 1+\frac{1}{2}\left[
1+x_{1}+x_{2}\left( 1-\frac{b_{3}}{2}\right) \eta \right] \left[ Z_{\text{s}%
}(\eta )-1\right] .  \label{11}
\end{equation}
Therefore, while Eq.\ (\ref{5}) is consistent with the exact property (\ref
{9}), Eq.\ (\ref{1}) violates it. In fact, the right-hand side of (\ref{7}),
with $\xi =x_{1}$, gives the deviation of Barrio and Solana's EOS from the
exact compressibility factor in the special case $\alpha \rightarrow 0$.

In summary, in this report we have performed a detailed comparison of the
EOS proposed by Barrio and Solana \cite{BS01} and the one introduced by 
 {Santos \textit{et al.}}
\cite{SYH99}. We find that both proposals lead to comparable accuracy when
the same EOS for the single component system is used and confirm that the
more accurate the $Z_{\text{s}}(\eta )$ the more accurate the resulting
compressibility factor for the binary mixture. In favor of  {Santos 
\textit{et al.}}'s proposal,
apart from its simpler form which also yields a very reasonable estimate of
the known third virial coefficient, is the fact that it is readily
extendible to the multicomponent case (including polydisperse mixtures) and 
complies with the exact limit in
which the small disks are point particles, while 
$Z_{\text{m}}^{\text{BS}}(\eta )$ does not share these assets.
\begin{acknowledgments}
The research of M.L.H. was supported in part by DGAPA-UNAM under Project
IN103100. S.B.Y. and A.S. acknowledge partial support from the Ministerio de
Ciencia y Tecnolog\'{\i}a (Spain) and FEDER through grant No.\ BFM2001-0718
\end{acknowledgments}

\end{document}